# Screening-Induced Phase Transitions in Core-Shell Ferroic Nanoparticles

Anna N. Morozovska[1*], Eugene A. Eliseev[2], Yulian M. Vysochanskii[3], Viktoria V. Khist[4], and Dean R. Evans[5†]

[1] Institute of Physics, National Academy of Sciences of Ukraine, 46, pr. Nauky, 03028 Kyiv, Ukraine

[2] Institute for Problems of Materials Science, National Academy of Sciences of Ukraine, Krjijanovskogo 3, 03142 Kyiv, Ukraine

[3] Institute of Solid State Physics and Chemistry, Uzhhorod University, 88000 Uzhhorod, Ukraine

[4] Igor Sikorsky Kyiv Polytechnic Institute, Kyiv, Ukraine

[5] Air Force Research Laboratory, Materials and Manufacturing Directorate, Wright-Patterson Air Force Base, Ohio, 45433, USA

**Abstract**

Using the Landau-Ginzburg-Devonshire approach, we study screening-induced phase transitions in core-shell ferroic nanoparticles for three different shapes: an oblate disk, a sphere, and a prolate needle. The nanoparticle is made of a ferroic $CuInP_2S_6$ core and covered by a "tunable" screening shell made of a phase-change material with a conductivity that varies as the material changes between semiconductor and metallic phases. We reveal a critical influence of the shell screening length on the phase transitions and spontaneous polarization of the nanoparticle core. Since the tunable screening shell allows the control of the polar state and phase diagrams of core-shell ferroic nanoparticles, the obtained results can be of particular interest for applications in nonvolatile memory cells.

---

[*] Corresponding author, e-mail: anna.n.morozovska@gmail.com

[†] Corresponding author: dean.evans@afrl.af.mil

# I. INTRODUCTION

The role of the surface state increases significantly with a decrease in the size of nanoscale ferroics often leading to size-induced phase transitions [1, 2], which can result in unusual polarization dynamics and morphology of polar domains related with the surface screening [3, 4]. The surface screening becomes especially important in recently discovered nanoscale Cu-based layered chalcogenides, $CuInP_2(S,Se)_6$ [5, 6, 7], which are uniaxial ferroics [8, 9, 10] with a possibility of the ferrielectricity and antiferrielectricity downscaling to the limit of a single layer [11, 12]. Ferrielectricity, the equivalent of ferrimagnetism, can be termed as an antiferroelectric order, but with a switchable spontaneous polarization created by two sublattices with spontaneous dipole moments that are antiparallel and different in magnitude [13, 14]. From a microscopic point of view $CuInP_2(S,Se)_6$ nanoparticles are ferrielectrics [13]; the macroscopic spontaneous polarization of $CuInP_2S_6$ ranges from 0.03 $C/m^2$ to 0.12 $C/m^2$ [15], while for $CuInP_2Se_6$ it is approximately 0.025 $C/m^2$ [16]. The spontaneous polarization occurs at ~(310 - 320) K for $CuInP_2S_6$ and at ~230 K for $CuInP_2Se_6$ [17, 18]. The $CuInP_2(S,Se)_6$ family reveals very unusual features: nonlinear dielectric response indicating that a spontaneous polarization may exist above the ferrielectric transition temperature [19], extremely large elastic nonlinearity in the direction perpendicular to the layers [20, 21], giant negative electrostriction, giant piezoelectricity, and dielectric tunability [22], electrostriction-induced piezoelectricity above the ferrielectric transition temperature [23], morphotropic phase transitions between the monoclinic and trigonal phases [24], anomalous "bright" domain walls with an enhanced local piezoelectric response [25, 26], giant flexoelectric response [12], and stress-induced phase transitions in $CuInP_2S_6$ ellipsoidal nanoparticles [27].

Despite the significant fundamental and practical interest in nanoscale $CuInP_2(S,Se)_6$, the influence of surface screening on their phase diagrams and polar properties are poorly studied. Here, very interesting effects can be expected in $CuInP_2(S,Se)_6$ nanoparticles covered by tunable shells; this expectation is based on our previous experience. Recently, we simulated very unusual polar states, such as labyrinthine patterns [28, 29], vortices with kernels [30, 31], twisted vortices [32], and meron-like flexons [33] in a spherical (or cylindrical) ferroelectric core covered with a "tunable" paraelectric shell placed in a soft polymer (or liquid medium). In particular, labyrinthine domains can exist in spherical core-shell nanoparticles with a uniaxial ferrielectric $CuInP_2S_6$ or ferroelectric $Sn_2P_2S_6$ core, and may have a quasi-infinite number of equal-energy states [28, 29]. Vortex states with a kernel may occur in spherical core-shell nanoparticles with a multiaxial ferroelectric $BaTiO_3$ or $BiFeO_3$ core, which possess a manifold degeneracy [30-32]. The great number of the equal-energy domain states may be the reason for a low-frequency negative susceptibility [34].

Using the four-well Landau-Ginzburg-Devonshire (**LGD**) thermodynamic potential reconstructed in Ref.[27], we study the screening-induced phase transitions in core-shell nanoparticles, whose shape varies, i.e., a prolate needle, an oblate disk, or a sphere. The nanoparticle core is made of CuInP$_2$S$_6$ (**CIPS**). The original part of this work briefly discusses basic expressions of the LGD approach (**Section II**), and considers in detail the screening-induced phase transitions and polarization changes in these core-shell nanoparticles (**Section III**). **Section IV** provides conclusions of the results.

## II. Problem formulation

A core-shell nanoellipsoid with semi-axes $R$ and $L$ is shown in **Fig. 1a**. The spontaneous polarization of the CIPS core, $P_3$, is directed along the ellipsoid semi-axis $L$. The macroscopic polarization $P_3$ is the total polarization of four possible polar-active sublattices $U_3^{(i)}$, $P_3 = \frac{e}{2}\left(U_3^{(1)} + U_3^{(2)} + U_3^{(3)} + U_3^{(4)}\right)$ (see **Fig. 1b** and Eq.(2a) in Ref.[26]), in which individual dipole moments of the four Cu and four In atoms are antiparallel and different in magnitude in the absence of external electric field and applied pressure. Polarizations created by individual Cu and In dipoles and resulting polarization $P_3$ are shown in **Fig. 1c**, where we use experimental data from Fig. 4a in Ref.[13]. The polarization $P_3$ is a polar long-range order parameter, which can be directly determined by ferroelectric measurements and piezoelectric response microscopy. The antipolar order parameter, $A_3$, is defined as one of three possible half-differences of the sublattices polarizations, e.g., $A_3 = \frac{e}{2}\left(U_3^{(1)} - U_3^{(2)} - U_3^{(3)} + U_3^{(4)}\right)$ (see Eq.(2b) in Ref.[26]). The antipolar order parameters cannot be directly measured in the abovementioned experiments. Indirectly, the nonlinear coupling between $A_3$ and $P_3$ increases the order of the thermodynamic potential, $G$ (e.g., Gibbs free energy), describing the polar behavior of CIPS (see **Appendix A** for details).

It has been shown [27] that CIPS free energy, $G(P_3)$, can have one potential well at $P_3 = 0$, three wells at $P_3 = 0$ and $P_3 = \pm P_{S1}$, two wells at $P_3 = \pm P_{S2}$, two plateaus at $P_{S2} \leq |P_3| \leq P_{S1}$, or four wells at $P_3 = \pm P_{S2}$ and $P_3 \approx \pm P_{S1}$; Gibbs free energy is dependent on the value of applied stress $\sigma$ and temperature $T$.

Typical dependences of the CIPS free energy density $g$ on polarization $P_3$ calculated for $\sigma < 0$ and $\sigma > 0$ for different values of $T$ are shown in **Fig. 1d.** The paraelectric-ferrielectric phase transition is of the second order for the tension of CIPS lattice (see the top plot in **Fig. 1d**, where $\sigma < 0$). The second order scenario is realized, because the distance between the sub-lattices increases and interaction between the layers becomes weaker in the tensile lattice. As a result, the long-range electrostatic interaction between the sub-lattice dipoles becomes weaker and so their antiparallel ordering in the four sub-lattices becomes favorable in order to minimize the electrostatic energy of dipole-dipole interactions (small black arrows are antiparallel in the right atomic structure of **Fig. 1b**). This is turn provides local antiferroelectric state with $P_3 = 0$ and $A_3 \neq 0$ can occur in a given unit cell, consisting of four sub-lattices (see e.g., three possible antiferroelectric orderings in Fig. 1b-d in Ref.[26]). In accordance with ab initio calculations and earlier results [27], the local antiferroelectric orderings (such as shown in the right **Fig. 1b**) mix with local ferrielectric ordering (such as shown in the left side of **Fig. 1b**). In result, a small spontaneous polarization, $P_3 = \pm P_{S2}$, arises continuously under the temperature decrease below $T_C$ (see the appearance and deepening of two relatively shallow potential wells in the top plot of **Fig. 1d**).

The paraelectric-ferrielectric phase transition is of the first order for the undeformed and compressed CIPS lattice (see the bottom plot in **Fig. 1d**, where $\sigma > 0$). The first order scenario exists, because the distance between the sub-lattices in the compressed lattice. As a result, the long-range electrostatic interaction between the sub-lattice dipoles becomes stronger and so their parallel ordering in the four sub-lattices becomes favorable in order to minimize the electrostatic energy of dipole-dipole interactions (small black arrows are parallel in the left atomic structure of **Fig. 1c**). This results in a large spontaneous polarization, $P_3 = \pm P_{S1}$, which arises sharply below $T_C$ (see the formation and deepening of two relatively deep potential wells separated by potential barrier in the bottom plot of **Fig. 1d**).

The core is covered by a "tunable" screening shell made of a phase-change material, whose electric conductivity varies as the material changes between semiconductor and metallic phases. The metallic state shell with a greater conductivity corresponds to a very small effective screening length, $\lambda < 0.05$ nm, which effectively screens the depolarization electric field, supports the spontaneous polarization of the core, and can prevent the domain splitting in the core. The semiconductor state shell with a relatively weaker conductivity corresponds to a larger screening length, $\lambda > 0.5$ nm, resulting in a decrease of the core polarization and a possible appearance of domain splitting in order to minimize the electrostatic energy of a depolarization field in the system. Note that any type of mechanical action (e.g., hydrostatic pressure, biaxial or uniaxial mechanical stress) can change the nanoparticle polar state due to the electrostriction effect [27].

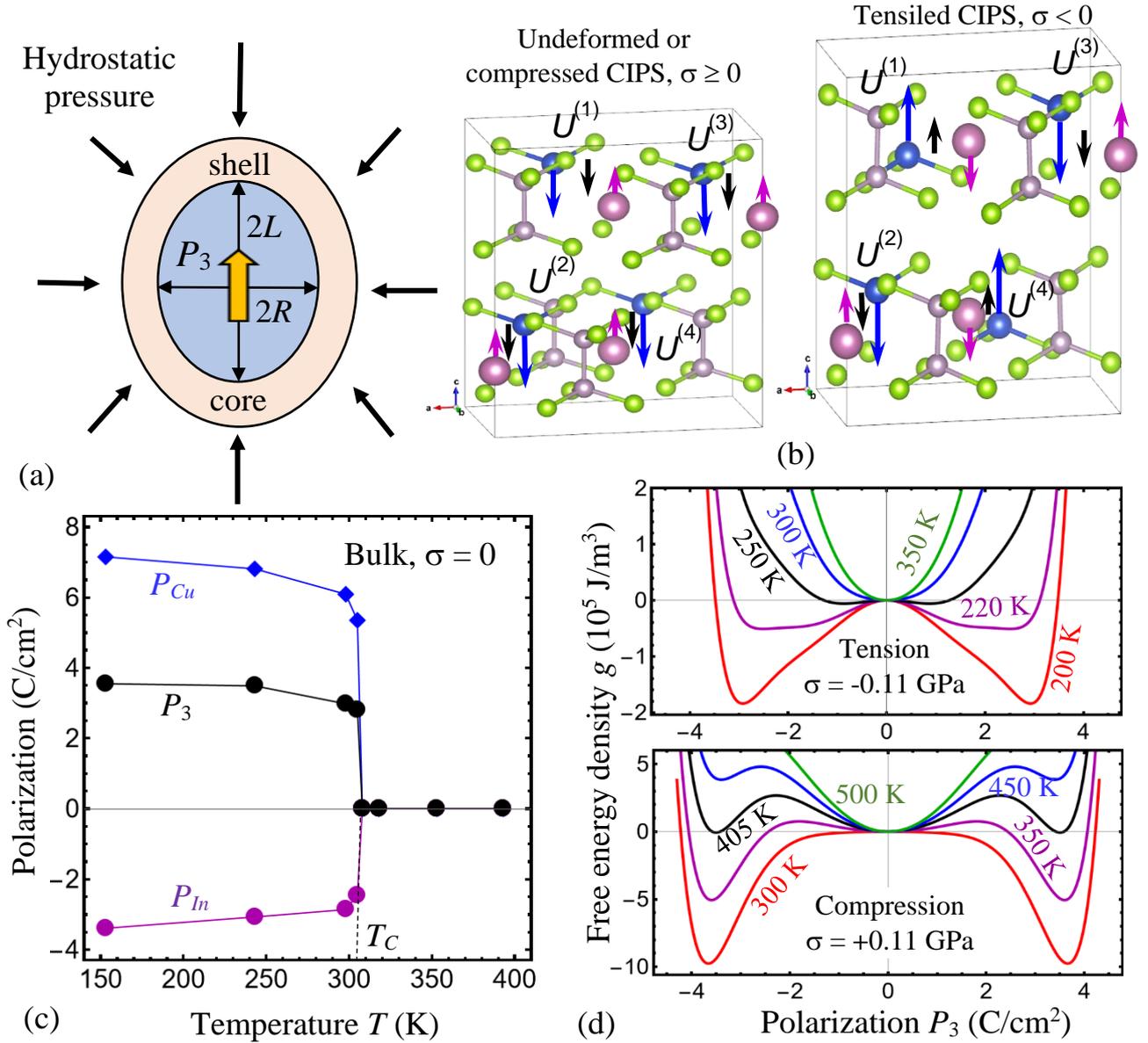

**FIGURE 1. (a)** Core-shell nanoellipsoid with semi-axes $R$ and $L$. The thick orange arrow shows the spontaneous polarization directed along the ellipsoidal axis $L$, thin black arrows illustrate the direction of the hydrostatic pressure application (compression is depicted in the schematic). **(b)** Atomic structure of CIPS [25], where the blue, large violet, small grey-violet, and yellow balls are Cu, In, P, and S atoms, respectively. Violet and blue arrows schematically illustrate the opposite dipole moments of In and Cu sub-lattices, respectively. Black arrows are resulting polar displacements $U^{(m)}$ in two ferrielectric states with large (left structure) and zero (right structure) total polarizations. The orientation of crystallographic axis a, b, and c is shown at the bottom of each structure. **(c)** Temperature dependence of polarizations created by individual Cu and In dipoles in the ferrielectric phase and the resulting net polarization $P_3$. Symbols are experimental data from Fig. 4a in Ref.[13], blue, black, and violet solid curves are theoretical fitting for a bulk CIPS at $\sigma = 0$. **(d)** Typical dependences of the free energy on polarization $P_3$ calculated for a bulk CIPS at negative (top plot) and positive pressures, $\sigma = -0.11$ GPa (top plot) and $\sigma = +0.11$ GPa (bottom plot). Red, violet, black, blue, and green curves correspond to different temperatures $T$ from 200 K to 500 K.

The density of the four-well LGD potential, $g_{LGD}$, which includes the Landau-Devonshire expansion in even powers of the polarization $P_3$ up to the eight power, $g_{LD}$, the Ginzburg gradient energy $g_G$, and the elastic and electrostriction energies, $g_{ES}$, has the form [27]:

$$g_{LGD} = g_{LD} + g_G + g_{ES}, \tag{1a}$$

$$g_{LD} = \frac{\alpha^*}{2} P_3^2 + \frac{\beta}{4} P_3^4 + \frac{\gamma}{6} P_3^6 + \frac{\delta}{8} P_3^8 - P_3 E_3, \tag{1b}$$

$$g_G = g_{33ij} \frac{\partial P_3}{\partial x_i} \frac{\partial P_3}{\partial x_j}, \tag{1c}$$

$$g_{ES} = -\frac{s_{ij}}{2} \sigma_i \sigma_j - Q_{i3} \sigma_i P_3^2 - Z_{i33} \sigma_i P_3^4 - W_{ij3} \sigma_i \sigma_j P_3^2. \tag{1d}$$

In accordance with LGD theory, the coefficients $\beta$, $\gamma$, and $\delta$ in Eq.(1b) are temperature independent. The values $\sigma_i$ denote diagonal components of a stress tensor in the Voigt notation, and a subscript $i = 1 - 6$. The values $Q_{i3}$, $Z_{i33}$, and $W_{ij3}$ denote the components of a single linear and two nonlinear electrostriction strain tensors, respectively [35, 36]. $E_3$ is an electric field component co-directed with the polarization $P_3$. The last term is the energy of a polarization gradient, where the strength and anisotropy are defined by the tensor $g_{33ij}$. In order to focus on the influence of external pressure, we neglect the surface tension and polarization gradient effects considered elsewhere [37, 38, 39].

In Eq.(1b) the temperature-, shape-, size-, and screening-dependent function $\alpha^*$ is introduced [27]:

$$\alpha^*(T, n_d, \Lambda) = \alpha(T) + \frac{n_d}{\varepsilon_0 [\varepsilon_b n_d + \varepsilon_s (1 - n_d) + n_d \Lambda]}. \tag{2}$$

Here the coefficient $\alpha$ depends linearly on the temperature $T$, $\alpha(T) = \alpha_T (T - T_C)$, where $T_C$ is the Curie temperature of the bulk ferrielectric. The derivation of Eq.(2) is given in Ref.[40]. The dimensionless parameter $n_d$ is the shape-dependent depolarization factor introduced as [41]:

$$n_d(\zeta) = \frac{1-\xi^2}{\xi^3} \left( \ln \sqrt{\frac{1+\xi}{1-\xi}} - \xi \right), \quad \xi(\zeta) = \sqrt{1-\zeta^2}, \quad \zeta = \frac{R}{L}. \tag{3}$$

Here $\xi$ is the eccentricity ratio of the ellipsoid with semi-axes $R$ and $L$, and $\zeta$ is the dimensionless shape factor. Parameters $\varepsilon_b$ and $\varepsilon_s$ are the background dielectric permittivity [42] of the ferrielectric core and the relative dielectric permittivity of the shell, respectively. The dimensionless screening factor $\Lambda$ is introduced as:

$$\Lambda = \frac{\lambda}{L}. \tag{4}$$

Let us consider nanoparticles with a small size, $L < 10$ nm. In this case, small screening factors, $\Lambda < 0.01$, correspond to a high screening degree provided by the conducting (i.e., metallic and semi-metallic) state of the phase-change tunable shell. Larger screening factors, $0.1 < \Lambda < 1$, correspond to a low screening degree provided by the semiconducting state of the tunable shell.

The values of $T_C$, $\alpha_T$, $\beta$, $\gamma$, $\delta$, $Q_{i3}$, and $Z_{i33}$ have been determined in Refs.[26, 27] from the fitting of temperature dependent experimental data for the dielectric permittivity [43, 44, 45], spontaneous polarization [13], and lattice constants [10] as a function of hydrostatic pressure. Elastic compliances $s_{ij}$ were estimated from ultrasound velocity measurements [21, 23, 46]. The details for determining the CIPS material parameters are given in the Supplement of Ref.[27]. Unfortunately, we did not find a full set of reliable experimental data for CIPS that would allow us to determine all nonlinear electrostriction coefficients $W_{ijk}$ with satisfactory accuracy. However, we managed to estimate the diagonal components, $W_{ii3}$, which are coupled with the hydrostatic pressure in Eq.(1d) using the experimental results in Ref.[47]. The gradient coefficients $g_{33ij}$ are not determined in Refs.[26, 27], but estimated from the width of domain walls. The CIPS parameters are listed in **Table I.**

**Table I.** LGD parameters for a bulk ferrielectric CuInP$_2$S$_6$.

| coefficient | value |
|---|---|
| $\varepsilon_b$ | 9 |
| $\alpha_T$ (C$^{-2}$·m J/K) | 1.64067×10$^7$ |
| $T_C$ (K) | 292.67 |
| $\beta$ (C$^{-4}$·m$^5$J) | 3.148×10$^{12}$ |
| $\gamma$ (C$^{-6}$·m$^9$J) | −1.0776×10$^{16}$ |
| $\delta$ (C$^{-8}$·m$^{13}$J) | 7.6318×10$^{18}$ |
| $Q_{i3}$ (C$^{-2}$·m$^4$) | $Q_{13} = 1.70136 - 0.00363\,T$, $Q_{23} = 1.13424 - 0.00242\,T$, $Q_{33} = -5.622 + 0.0105\,T$ |
| $Z_{i33}$ (C$^{-4}$·m$^8$) | $Z_{133} = -2059.65 + 0.8\,T$, $Z_{233} = -1211.26 + 0.45\,T$, $Z_{333} = 1381.37 - 12\,T$ |
| $W_{ij3}$ (C$^{-2}$·m$^4$ Pa$^{-1}$) | $W_{113} \approx W_{223} \approx W_{333} \cong -2 \times 10^{-12}$ |
| $s_{ij}$ (Pa$^{-1}$) | $s_{11} = 1.510 \times 10^{-11}$, $s_{12} = 0.183 \times 10^{-11}$ |
| $g_{33ij}$ (J m$^3$/C$^2$) | $g \cong (0.5 - 2.0) \times 10^{-9}$ |

### III. Screening-induced phase transitions in core-shell nanoparticles

Here we account for the effect of the hydrostatic pressure $\sigma$ on the spontaneous polarization and phase diagram of the core-shell nanoparticles. The phase diagram in **Fig. 2a** illustrates the typical influence of the hydrostatic pressure on the polar state of an ellipsoidal CIPS nanoparticle. It contains a large dark-violet region representing the paraelectric phase (**PE**) that expands towards higher temperatures, and a region of the ferrielectric (**FI**) phase, which has two states. A smaller reddish region represents the ferrielectric (**FI**) state 1 (abbreviated as "**FI1**") with a relatively large polarization $P_3$ that corresponds to the compression of the particle ($\sigma > 0$), and a small bluish region of the ferrielectric state 2 (abbreviated as "**FI2**") with a small $P_3$ corresponds to the particle expansion under tension ($\sigma < 0$). The phase diagram has a critical end point (**CEP**), marked by a black triangle and located at $\sigma_{CEP} \approx -0.05$ GPa and $T_{CEP} \approx 180$ K; and a bicritical end point (**BEP**) [27, 48], marked by a white circle and located at $\sigma_{BEP} \approx -0.11$ GPa and $T_{BEP} \approx 100$ K. The dependences of the free energy $g_{LGD}$ on the polarization $P_3$ calculated for different pressures and temperatures are shown in **Fig. 2b-d.**

Phase diagrams and the FI-PE transition order depend on pressure due to the electrostriction coupling terms, $-Q_{i3}\sigma_i P_3^2 - Z_{i33}\sigma_i P_3^4 - W_{ij3}\sigma_i\sigma_j P_3^2$, in Eq.(1d). The coupling changes the coefficients $\frac{\alpha^*}{2}$ and $\frac{\beta}{4}$ in the Landau-Devonshire energy Eq.(1b) to $\left(\frac{\alpha^*}{2} - Q_{i3}\sigma_i - W_{ij3}\sigma_i\sigma_j\right)$ and $\left(\frac{\beta}{4} - Z_{i33}\sigma_i\right)$, respectively. Since $\beta > 0$, $\gamma < 0$, and $\delta > 0$, the pressure $\sigma_{cr}$ and temperature $T_{cr}$, which correspond to the BEP, are determined from the conditions [27]:

$$(Z_{133} + Z_{233} + Z_{333})\sigma_{cr} = \frac{1}{4}\left(\beta - \frac{\gamma^2}{3\delta}\right), \tag{5a}$$

$$\alpha_T(T_{cr} - T_C) = 2\sigma_{cr}(Z_{133} + Z_{233} + Z_{333}) + 2(W_{113} + W_{223} + W_{333})\sigma_{cr}^2 + \frac{\gamma^3}{27\delta^2}, \tag{5b}$$

where $Q_{i3}$ and $Z_{i33}$ linearly depends on $T$. Due to the terms $\frac{\gamma}{6}P_3^6 + \frac{\delta}{8}P_3^8$, negative nonlinear electrostriction couplings $Z_{i33} < 0$ and $W_{ij3} < 0$, and the "inverted" signs of the linear electrostriction coupling $Q_{33} < 0$, $Q_{23} > 0$ and $Q_{13} > 0$ for CIPS, the pressure effect on the phase diagram is complex and unusual [27].

At $\sigma < \sigma_{CEP}$ the second order FI-PE transition occurs through the FI2 state with a diffuse boundary (see the green diffuse boundary in **Fig. 2a** and the dependences of $g_{LGD}$ on $P_3$ in **Fig. 2b**). The diffuseness means that the spontaneous polarization changes continuously from zero to nonzero values inside the FI2 state. The FI-PE second order transition line at CEP meets the FI-PE first order transition line. At $\sigma < \sigma_{CEP}$ the first order transition line is in the FI phase, and a sharp increase of the $P_3$ value is observed at this line. The magnitude of this increase from a small $P_3$ value in the FI2 state to a large $P_3$ value in the FI1 state decreases with an increase of tensile stress, until the stress and temperature values approach the BEP (note, BEP is characterized by $\sigma_{BEP} \approx -0.11$ Gpa and $T_{BEP} \approx 100$ K). At $\sigma < \sigma_{BEP}$, a second order transition occurs, resutling in a continuous increase of the $P_3$ when moving from the FI2 state to the FI1 state.

At $\sigma > \sigma_{CEP}$ the first order FI-PE transition appears in the region of coexisting PE phase and FI1 state (abbreviated as "**FI1+PE**"), where the FI1 state is stable and the PE phase is metastable. The spontaneous polarization has a steep drop at the PE phase boundary (see the sharp reddish-violet boundary between the PE and FI1+PE regions in **Fig. 2a** and the dependences of $g_{LGD}$ on $P_3$ in **Fig. 2c-d**).

The coexistence of the FI and PE phases occurs around the first order transition and the coexisting region expands with a positive pressure increase (see Ref. [27] for details). The white solid and dashed curves position is not seen at the color scheme, because the numerical algorithm calculates the value of the spontaneous polarization only, and cannot estimate the ratio of the FI1 to PE clusters. The FI1+PE region in **Fig. 2a** is the color map of the spontaneous polarization, but not the average polarization of FI1 and PE clusters in the region. The white solid curve, which separates the FI1 and FI1+PE regions, corresponds to a complete disappearance of the metastable potential well at $P_3 = 0$, and is always inside the FI1 region. The white dashed curve corresponds to a complete disappearance of the metastable potential well at $P_3 = \pm P_S$, and is always inside the PE region at $\sigma > \sigma_{CEP}$. Note that both wells, $P_3 = 0$ and $P_3 = \pm P_S$, transform continuously into the one flat well in the BEP (see **Fig. 2b** and **2c**).

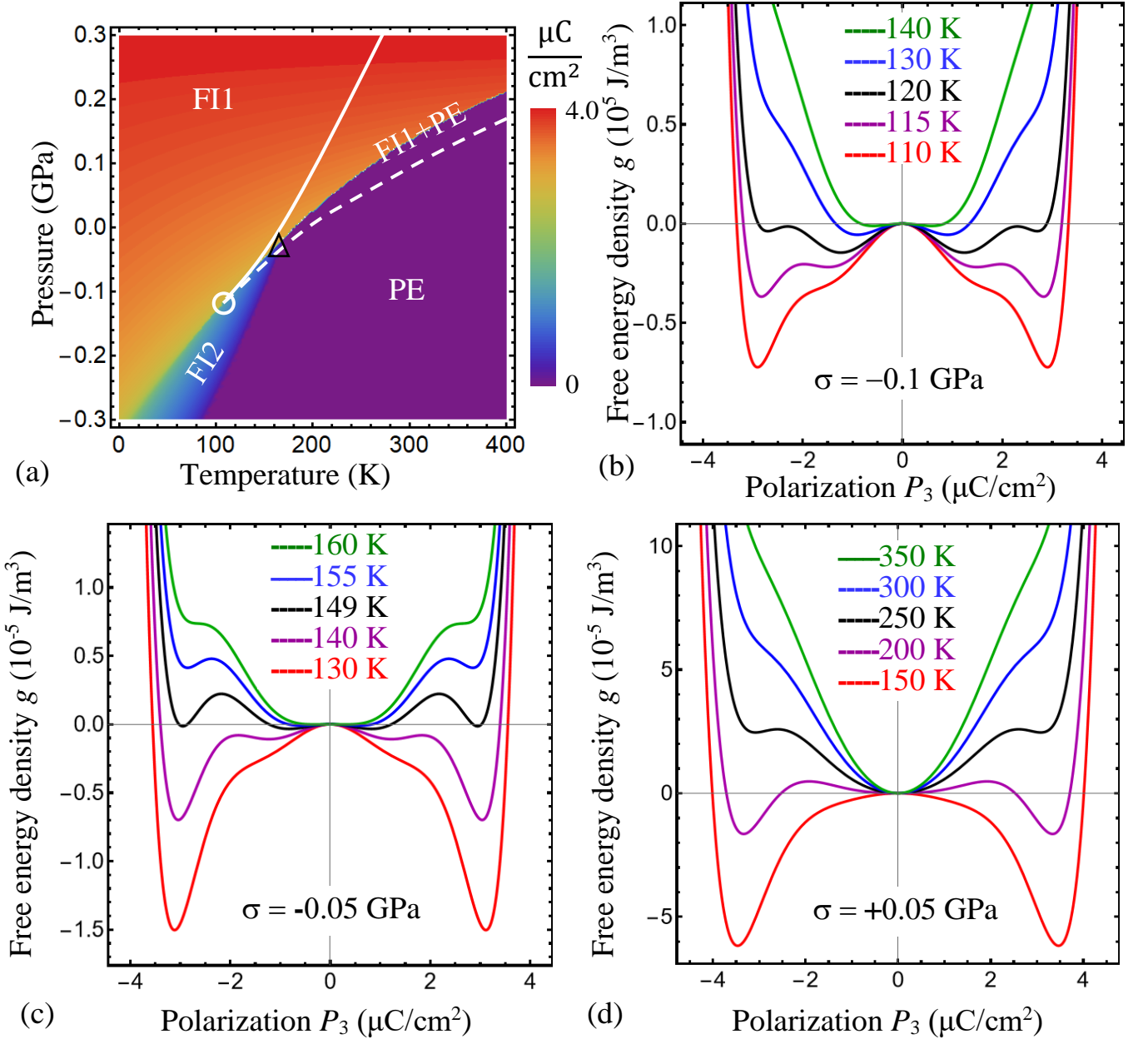

**FIGURE 2.** (a) The dependence of the spontaneous polarization $P_3$ on temperature and pressure calculated for the stressed CIPS nanoellipsoid with semi-axes $R = 10$ nm and $L = 20$ nm. White solid and dashed curves are the boundaries of the PE and FI phases absolute instability, respectively. A critical end point is marked by a black triangle, and a bicritical end point is marked by a white circle (see explanation in the text). The dependences of the free energy on polarization $P_3$ calculated for the pressures $\sigma = -0.1$ GPa **(b)**, $\sigma = -0.05$ GPa **(c)**, and $\sigma = +0.05$ GPa **(d)**. Red, violet, black, blue, and green curves correspond to different temperatures $T$ from 110 K to 350 K (see legends). The effective screening length $\lambda = 0.5$ nm and permittivity $\varepsilon_e = 2$, other CIPS parameters are listed in **Table I**.

As it follows from Eqs.(1b) and (1d), the critical temperature $T_{cr}$ of the PE phase instability is determined from the equation $\frac{1}{2}\alpha^*(T_{cr}, \zeta, \Lambda) - \sigma_i Q_{i3} - W_{ij3}\sigma_i\sigma_j = 0$, where the coefficient $\alpha^*(T_{cr}, \zeta, \Lambda)$ is given by Eq.(2). As illustrated below, the screening factor $\Lambda$, the shape factor $\zeta$, and the elastic stresses $\sigma_i$ can strongly change the value of $T_{cr}$, and, consequently, can influence the polar properties of the core at the working temperature. Since different shapes of core-shell nanoparticles are of interest for fundamental research and applications, the influence of the screening factor $\Lambda$ on their phase diagrams and polar properties is illustrated in **Fig. 3-6** for nanoparticles with various shape factors $\zeta$ under different pressures $\sigma$.

Phase diagrams of the core-shell CIPS nanoparticles as a function of both shape ($\zeta$) and screening ($\Lambda$) factors are shown in **Fig. 3** for several temperatures over the range (100 – 300) K. The diagrams **(a) - (f)** correspond to different values of hydrostatic pressure $\sigma$ varying in the range $-300$ MPa $\leq \sigma \leq$ 200 MPa. Note that the changes between the diagrams **(a)-(f)** are pressure-induced. The absence of red curves in the diagrams **(c) - (f)** means that the nanoparticle is in the PE phase at 300 K and $\sigma \leq 0$; the nanoparticle cannot be in the FI phase, because $T_{cr} < 0$ K for the negative pressures in the considered ranges of screening and shape factors (simply meaning that there is no FI-PE phase transition at $T \geq 0$ K). Similarly, the absence of a magenta curve in the diagram **(f)** means that the nanoparticle is in the PE phase at 250 K and $\sigma \leq -300$ MPa. At the temperatures below 200 K, the core-shell nanoparticles can be in the FE phase for $-300$ MPa $\leq \sigma \leq$ 200 MPa (see blue, green, and black curves in the diagrams **(a) - (f)**).

The common feature of the diagrams, shown in **Fig. 3**, is that the FI-PE transition temperature and the area of the FI phase increases for the case of compression, $\sigma > 0$, and decreases for the case of tension, $\sigma < 0$ (compare the position of red, magenta, and blue curves in the diagrams **(a) - (f)**). This trend is opposite to the situation observed for many uniaxial and multiaxial perovskite nanoparticles, where the FI-PE transition temperature and the area of the FI phase increases for negative hydrostatic pressure. The origin of this difference is the negative nonlinear electrostriction couplings $Z_{i33} < 0$ and $W_{ij33} < 0$, and the "inverted" signs of the linear electrostriction coupling $Q_{33} < 0$, $Q_{23} > 0$ and $Q_{13} > 0$ (see **Table SI**).

The PE phase, located in the top right part of the diagrams, enlarges its area with an increase in $\Lambda$ and/or $\zeta$, meaning that the phase stability increases with an increase in the screening factor (e.g., for semiconducting shells with $\Lambda > 0.1$), as well as with a decrease of the core length in the polarization direction (e.g., for spherical or oblate ellipsoidal cores with $\zeta \geq 1$). The FI phase, located in the bottom left part of the diagrams, enlarges its area with a decrease in $\Lambda$ and/or $\zeta$, meaning that the phase stability increases with a decrease in the screening factor (e.g., for conducting shells with $\Lambda < 0.01$), as well as with an increase of the core length in the polarization direction (e.g., for prolate ellipsoidal cores with $\zeta < 1$). As anticipated, the area of the PE phase decreases and the area of the FI phase increases with a decrease in temperature (compare the position of different curves in the diagrams).

Another common feature of great importance in the diagrams shown in **Fig. 3**, is their strong dependence on the screening factor $\Lambda$. Namely, the curves of the FI-PE phase transition have nearly vertical regions, which indicate a screening-induced nature of the transition. The transition occurs at a critical value of the screening length, $\Lambda = \Lambda_{cr}(\zeta, T, \sigma)$, which depends on the temperature and pressure. The screening-induced FI-PE transition is caused by the depolarization energy contribution, whose strength is determined by the $\Lambda$ value. Thus, the tunable screening shell is responsible for inducing, maintaining, or destroying the polar state in the nanoparticle core.

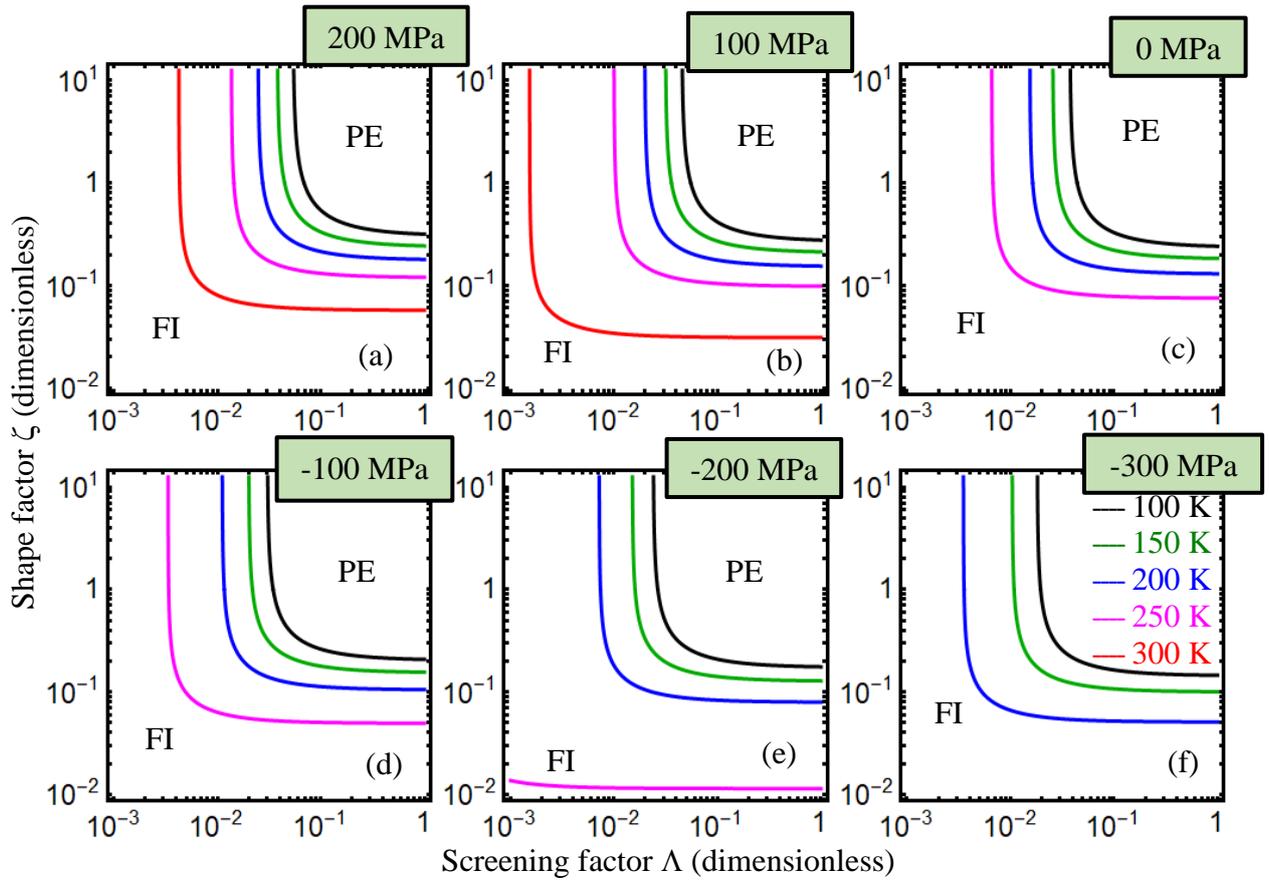

**Figure 3**. Phase diagrams of the core-shell CIPS nanoparticles and their dependence on the screening factor $\Lambda$ and shape factor $\zeta$. The diagrams are calculated for different temperatures $T = 100, 150, 200, 250,$ and $300$ K

(black, green, blue, magenta, and red curves, respectively), and hydrostatic pressures $\sigma = 200$ MPa **(a)**, 100 MPa **(b)**, 0 **(c)**, 100 MPa **(d)**, -200 MPa **(e)**, and -300 MPa **(f)**.

The dependence of the spontaneous polarization $P_3$ on both temperature $T$ and screening factor $\Lambda$, calculated for core-shell CIPS nanodisks (shape factor $\zeta = 100$), nanospheres (shape factor $\zeta = 1$), and nanoneedles (shape factor $\zeta = 0.1$), is shown in **Figs. 4**, **5,** and **6**, respectively. Color maps **(a) - (f)** correspond to different values of hydrostatic pressure $\sigma$ varying in the range $-300$ MPa $\leq \sigma \leq$ 200 MPa. The polarization vector is perpendicular to the disk surface in **Fig. 4**, and is pointed along the needle's long axis in **Fig. 6.**

The maps **(a)-(c)**, calculated for positive and zero pressures $\sigma$, contain a large reddish FI1 region with a relatively large polarization $P_3$ that has a sharp boundary with the violet PE region with zero $P_3$. The sharpness of the FI-PE boundary corresponds to a first order phase transition. The area of the FI1 state with a large $P_3$ decreases with a decrease in $\sigma$ [compare the maps **(a)-(c)**]. The maps **(d)-(f)**, calculated for negative pressures $\sigma$, contain significantly smaller FI1 regions of a moderate polarization $P_3$ that has a diffuse boundary with the FI2 region of a small $P_3$. The FI region also borders with the PE region of zero $P_3$. The diffuseness of the FI1-FI2 and FI2-PE boundaries corresponds to a second order phase transition. The diffuse area of the FI2 state significantly enlarges with the increase in the magnitude of negative pressure [compare the maps **(d)-(f)** for $\sigma < 0$].

The area of the FI1 state varies significantly as a function of the nanoparticle shape. This area is smallest for nanodisks (see **Fig. 4**), becomes very slightly larger for nanospheres (for which the polarization magnitude is greater), especially for $\sigma < 0$ (see **Fig. 5**), and significantly larger for nanoneedles (see **Fig. 6**), in this latter case it can fill the entire map for $\sigma > 150$ MPa [see **Fig. 6(a)**]. The shape-dependent changes of the FI1 state area and the corresponding changes of polarization are caused by the depolarization field contribution, which is proportional to the depolarization factor [1, 2]. The depolarization factor $n_d$, given by Eq.(3), is greatest for nanodisks with $\zeta \gg 1$ (with the polarization vector perpendicular to the disk surface), moderate for nanospheres, and decreases as $\frac{1}{\zeta}$ for nanoneedles with $\zeta \ll 1$ (with the polarization vector pointed along the needle long axis) (see **Appendix B**). The relative area of the diffuse FI2 state with small $P_3$, which exists for negative $\sigma$, is approximately the same for nanodisks, nanospheres, and nanoneedles [compare e.g., **Figs. 4(f), 5(f),** and **6(f)**]. The area of the PE phase without a spontaneous polarization is the largest for nanodisks (see **Fig. 4**), slightly less for nanospheres (see **Fig. 5**), and significantly smaller for the nanoneedles (see **Fig. 6**), where for the latter case is absent if $\sigma > 150$ MPa [see **Fig. 6(a)**]. Note that the shape of the FI-PE boundary (either a sharp, first order transition for $\sigma \geq 0$ or a diffuse, second order transition for $\sigma < 0$) is similar for the nanodisks and nanospheres, being close to a parabolic curve with a rather small slope followed by a nearly vertical drop for $\sigma \leq 100$ MPa, whereas the shape of the boundary for the nanoneedles is close to a meander with two rather steep slopes. This drastic difference originated from the specific dependence of the depolarization factor $n_d$ on the shape factor $\zeta$.

The common feature of the spontaneous polarization color maps, shown in **Figs. 4-6**, is their significant dependence on the screening factor $\Lambda$. The screening controls the polarization value, especially in the FI1 state with large $P_3$ values, whose sharp boundary strongly depends on $\Lambda$ and weakly depends on the temperature $T$. The screening role becomes a little weaker with the decrease in $P_3$, e.g., in the FI2 state with a small $P_3$ whose diffuse boundaries depend on $T$ and $\sigma$. The impact of the screening factor on $P_3$ is caused by the depolarization field, where the magnitude of the screening is proportional to the $\Lambda$ value; therefore, the tunable screening shell is responsible for inducing, maintaining, or destroying the spontaneous polarization of the nanoparticle core.

An important note to make is that phase diagrams and transition orders depend on pressure for many ferroics. The physical origin of the dependence is the electrostriction coupling. The order changes if the coupling renormalizes higher coefficients in LGD free energy. Since CIPS provides a negative nonlinear electrostriction, unlike most other ferroelectric materials, this condition (i.e., negative components $Z_{i33} < 0$ and $W_{ij3} < 0$) gives us a second order FI-PE transition for the case of negative pressures.

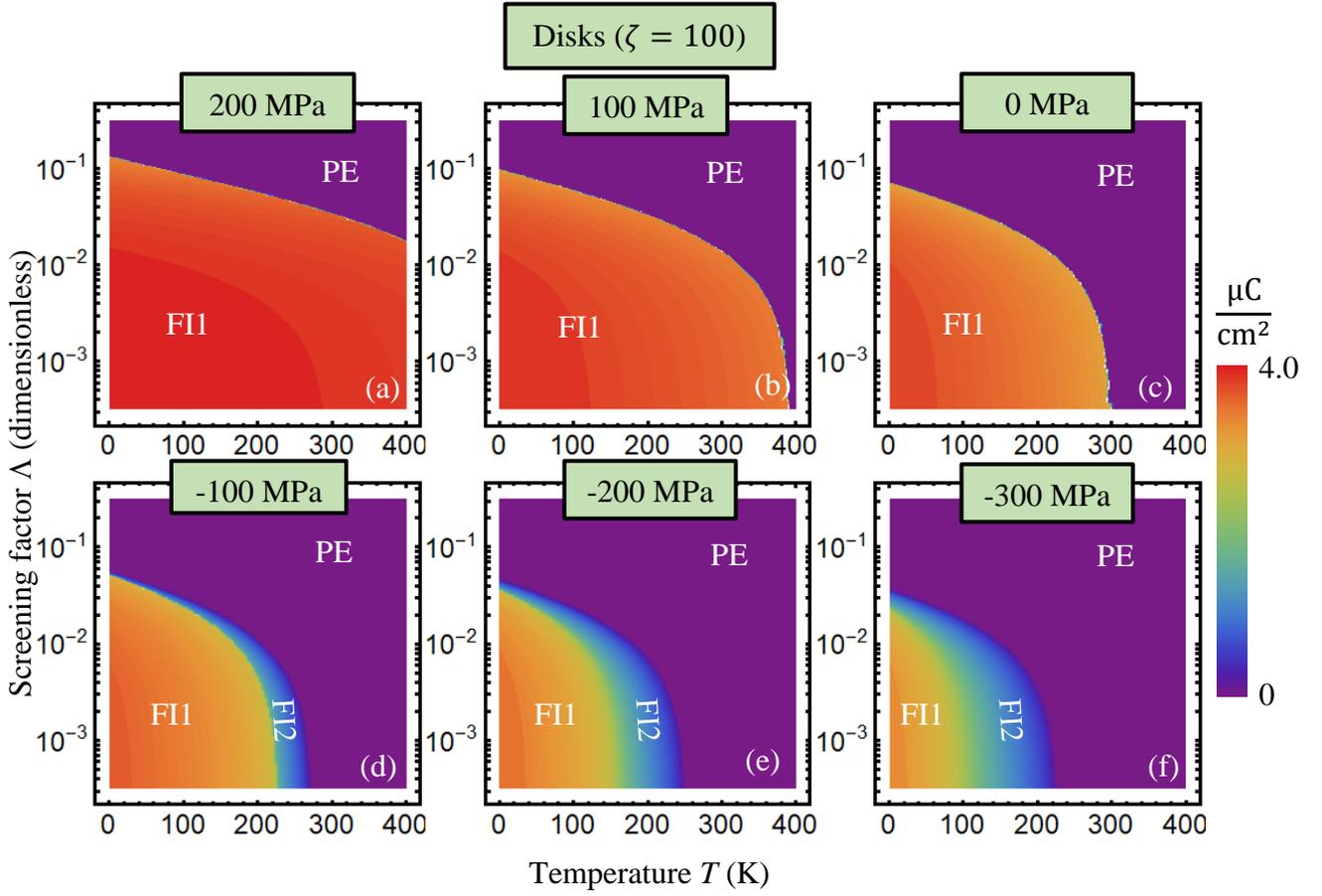

**Figure 4** The dependence of the spontaneous polarization $P_3$ on the temperature $T$ and screening factor $\Lambda$ calculated for core-shell nanodisks (shape factor $\zeta = 100$) at different hydrostatic pressures $\sigma = 200$ MPa **(a)**, 100 MPa **(b)**, 0 **(c)**, -100 MPa **(d)**, -200 MPa **(e)**, and -300 MPa **(f)**. The polarization vector is perpendicular to the disk surface.

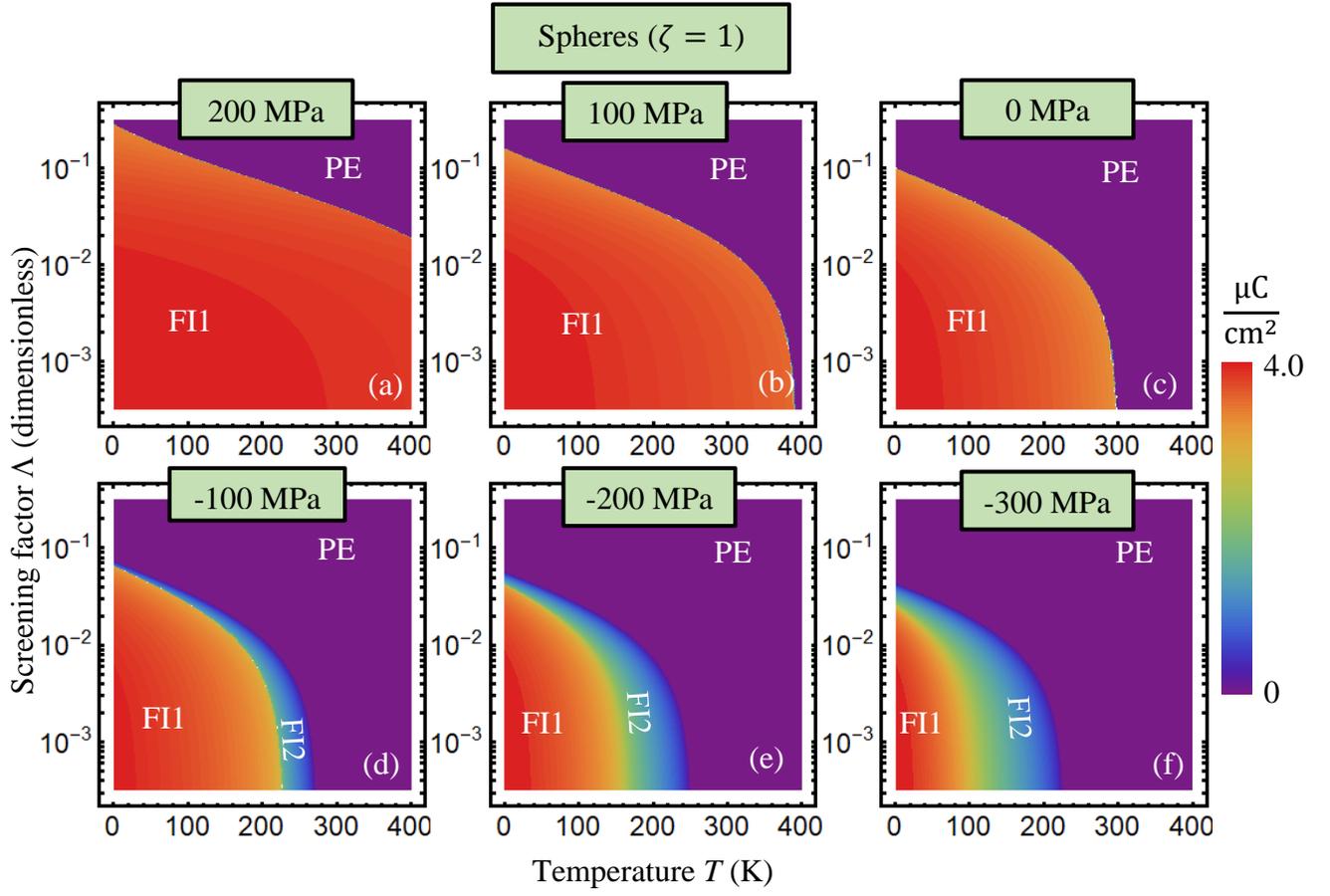

**Figure 5**. The dependence of the spontaneous polarization $P_3$ on the temperature $T$ and screening factor $\Lambda$ calculated for core-shell nanospheres (shape factor $\zeta = 1$) at different hydrostatic pressures: $\sigma = $ 200 MPa (**a**), 100 MPa (**b**), 0 (**c**), -100 MPa (**d**), -200 MPa (**e**), and -300 MPa (**f**).

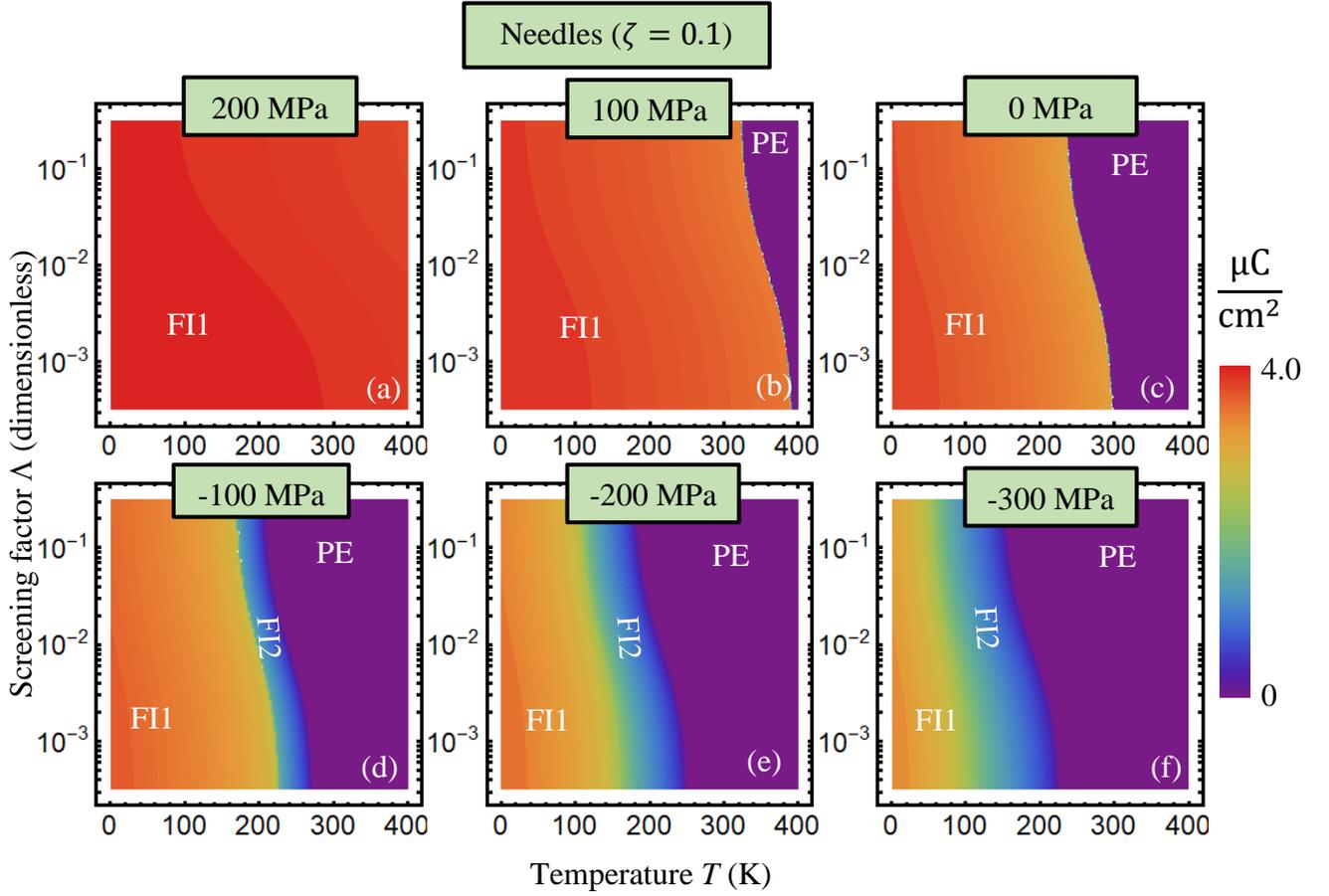

**Figure 6**. The dependence of the spontaneous polarization $P_3$ on the temperature $T$ and screening factor $\Lambda$ calculated for core-shell nanoneedles (shape factor $\zeta = 0.1$) at different hydrostatic pressures $\sigma = 200$ MPa **(a)**, 100 MPa **(b)**, 0 **(c)**, -100 MPa **(d)**, -200 MPa **(e)**, and -300 MPa **(f)**. The polarization vector is along the needle axis.

**Conclusions**

Using the LGD approach, we study screening-induced phase transitions in core-shell nanoparticles as a function of the particle shape: an oblate disk, a sphere, and a prolate needle. The nanoparticle core is made of a ferroic CIPS and covered by a tunable screening shell made of a phase-change material whose conductivity varies between semiconductor and metallic states. We revealed a very strong influence of the screening shell on the phase diagrams and polar properties of the nanoparticles.

In particular, the curves of FI-PE phase transition depend strongly on the screening factor $\Lambda$, which indicates a screening-induced nature of the transition. The transition occurs at some critical value of the screening length, which depends on the nanoparticle shape, temperature, and pressure. An important common feature is that phase transition order depends on pressure due to the electrostriction coupling. The order changes if the coupling renormalizes higher coefficients in LGD free energy. Since CIPS provides anegative nonlinear electrostriction, unlike most other ferroelectric materials, this condition (i.e., negative components $Z_{i33}$ and $W_{ij3}$) gives us a second order FI-PE transition for the case of negative pressures $\sigma < \sigma_{CEP}$.

The screening-induced FI-PE transition is caused by the depolarization energy contribution, where its strength is determined by the screening length. The FI phase has two states, FI1 and FI2, whose properties are screening-sensitive. The screening also controls the spontaneous polarization $P_3$, especially in the FI1 state with a large $P_3$, where the sharp boundary strongly depends on $\Lambda$. The screening effect becomes slightly weaker in the FI2 state with a small $P_3$, where its diffuse boundaries strongly depend on the temperature and pressure. The screening impact on $P_3$ is caused by the depolarization field, where its strength is proportional to the $\Lambda$ value.

Assuming that the ferrielectric FI1 and FI2 states, predicted theoretically, exist in real CIPS nanoparticles, they should coexist when their potential energies differ in value by an amount less than the thermal energy. Since the potential barrier between the close-energy FI1 and FI2 states can exist (e.g., at the temperatures between the black and dark-violet curves in **Fig. 2b**), the regions with the small and large $P_3$ can be randomly mixed in the nanoparticle core. The "mixture" can be interpreted as the spatially-inhomogeneous quasi-random polar phase. The switchable energy-degenerated FI1 and FI2 states can be promising candidates for multi-bits and high-T qubits. Since the tunable screening shell is responsible for inducing, maintaining, or destroying the polar state in the nanoparticle core, the results obtained in this study can be of particular interest for core-shell nanoparticles applications in nonvolatile memory cells.

**Methods.** Numerical results presented in the work are visualized using Mathematica 12.2 [49].

**Acknowledgements.** A.N.M. acknowledges EOARD project 9IOE063 and related STCU partner project P751a. E.A.E. acknowledges CNMS2021-B-00843 "Effect of surface ionic screening on polarization reversal scenario in antiferroelectric thin films: analytical theory, machine learning, PFM and cKPFM experiments".

**Authors' contribution.** Research ideas belongs to A.N.M. and D.R.E. A.N.M. formulated the problem, performed analytical calculations, analyzed results and wrote the manuscript draft. E.A.E. and V.V.K. wrote codes and prepared figures. Yu.M.V. and D.R.E. worked on the results explanation and manuscript improvement. All co-authors discussed the obtained results.

## Appendix A

A conventional expansion of a ferrielectric Landau-Devonshire thermodynamic potential contains even powers (2, 4, and 6) of the polar and antipolar order parameters, $P_3$ and $A_3$, and the biquadratic coupling between them. Due to the need to describe the first order phase transitions, we cannot cut the expansion on the fourth powers of $P_3$ and $A_3$; therefore, the expansion includes the sixth powers of $P_3$ and $A_3$. The corresponding expansion is:

$$G[P_3, A_3] = \frac{a}{2}P_3^2 + \frac{b}{4}P_3^4 + \frac{c}{6}P_3^6 - P_3 E_3 + \frac{d}{2}A_3^2 P_3^2 + \frac{f}{2}A_3^2 + \frac{g}{4}A_3^4 + \frac{h}{6}A_3^6. \tag{A.1}$$

Here $a$, $b$, $c$, $d$, $f$, $g$, and $h$ are expansion coefficients, and $E_3$ is an electric field. Minimization of $G[P_3, A_3]$ with respect to $A_3$ yealds:

$$\frac{\partial G}{\partial A_3} = A_3(dP_3^2 + f) + gA_3^3 + hA_3^5. \tag{A.2}$$

The solutions of equation $\frac{\partial G}{\partial A_3} = 0$ are $A_3 = 0$ and $A_3^2 = \frac{-g \pm \sqrt{g^2 - 4fh - 4dhP_3^2}}{2h}$. Substituting the nonzero solution for $A_3^2$ in Eq.(A.1), after algebraic transformations, and serial expansion on $P_3^2$, and further omitting highest order expansion terms proportional to $P_3^{10}$ and $P_3^{12}$, we obtain:

$$g_{LD}[P_3, A_3] = \frac{\alpha}{2}P_3^2 + \frac{\beta}{4}P_3^4 + \frac{\gamma}{6}P_3^6 + \frac{\delta}{8}P_3^8 - P_3 E_3, \tag{A.3}$$

where $\alpha \sim a$, $\beta \sim b$, $\gamma \sim c$, and $\delta \sim \frac{d^4 h^2}{g^5}$. So we can conclude that the biquadratic coupling term between the order parameters, $\frac{d}{2}A_3^2 P_3^2$, induces the term $\frac{\delta}{8}P_3^8$ in Eq.(A.3).

**Appendix B**

Most of the essential changes of the depolarization factor $n_d$ appear when the shape factor changes from very small values ($\zeta = \frac{R}{L} \ll 1$, needles) to larger values ($\zeta = 1 - 3$, sphere and sphere-like) (see **Fig. S1**). There is a saturation curve at $\zeta > 5$, when $n_d$ becomes close to unity and weakly independent on $\zeta$ for disk-like nanoparticles. The depolarization factor controls the transition temperature in accordance with Eq.(2), this means that for oblate nanoparticles (i.e., for disk-like) their shape becomes less important than other factors, such as temperature, pressure, and, of course screening length $\Lambda$. Note that the screening length $\Lambda$ is included in the depolarization factor, $\frac{n_d}{\varepsilon_0[\varepsilon_b n_d + \varepsilon_e(1-n_d) + n_d \Lambda]}$:

$$\alpha^*(T, n_d, \Lambda) = \alpha(T) + \frac{n_d}{\varepsilon_0[\varepsilon_b n_d + \varepsilon_e(1-n_d) + n_d \Lambda]}. \tag{B.1}$$

The depolarization factor controls the critical temperature in accordance with equation, $\frac{1}{2}\alpha^*(T_{cr}, n_d, \Lambda) - \sigma_i Q_{i3} - W_{ij3}\sigma_i\sigma_j = 0$, when $n_d$ is very small (for needles) $n_d \Lambda$ is also small, and so the phase boundaries in **Fig. 6** becomes more vertical, meaning that they are much more temperature-dependent than the case shown in **Fig. 4**.

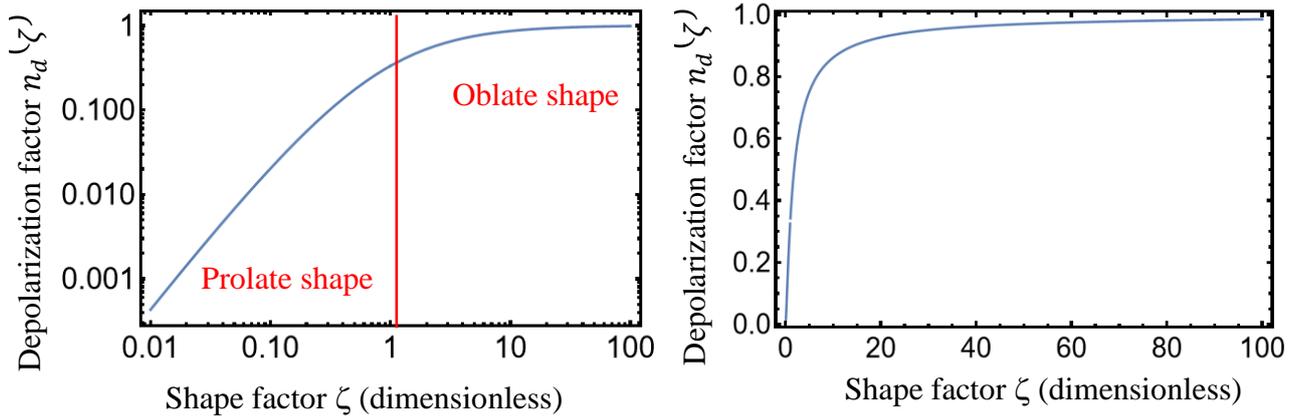

**Figure S1.** The dependence of the depolarization factor $n_d(\zeta)$ on the shape factor $\zeta = \frac{R}{L}$ plotted in log-log scale (left) and linear scale (right).

**References**

[1] A. N. Morozovska, E. A. Eliseev, M. D. Glinchuk. Ferroelectricity enhancement in confined nanorods : Direct variational method. Phys. Rev. B **73**, 214106 (2006).

[2] A. N. Morozovska, M. D. Glinchuk, E. A. Eliseev. Phase transitions induced by confinement of ferroic nanoparticles. Phys. Rev. B **76**, 014102 (2007).

[3] S. V. Kalinin, Y. Kim, D. Fong, and A. N. Morozovska, Surface-screening mechanisms in ferroelectric thin films and their effect on polarization dynamics and domain structures. Rep. Prog. Phys. **81**, 036502 (2018).

[4] A. N. Morozovska, E. A. Eliseev, S. V. Kalinin, and R. Hertel. Flexo-Sensitive Polarization Vortices in Thin Ferroelectric Films. Phys. Rev. B **104**, 085420 (2021).

[5] X. Bourdon, V. Maisonneuve, V.B. Cajipe, C. Payen, and J.E. Fischer. Copper sublattice ordering in layered $CuMP_2Se_6$ (M=In, Cr). J. All. Comp. **283**, 122 (1999).

[6] A. Belianinov, Q. He, A. Dziaugys, P. Maksymovych, E. Eliseev, A. Borisevich, A. Morozovska, J. Banys, Y. Vysochanskii, and S. V. Kalinin, $CuInP_2S_6$ Room Temperature Layered Ferroelectric. Nano Lett. **15**, 3808 (2015).

[7] M. A. Susner, M. Chyasnavichyus, M. A. McGuire, P. Ganesh, and P. Maksymovych. Metal Thio- and Selenophosphates as Multifunctional van der Waals Layered Materials. Advanced Materials **29**, 1602852 (2017).

[8] M. Wu, and P. Jena, The rise of two-dimensional van der Waals ferroelectrics. Wiley Interdisciplinary Reviews: Computational Molecular Science **8**, e1365 (2018).

[9] F. Liu, L. You, K. L. Seyler, X. Li, P. Yu, J. Lin, X. Wang, J. Zhou, H. Wang, H. He, S.T. Pantelides, W. Zhou, P. Sharma, X. Xu, P.M. Ajayan, J. Wang and Z. Liu, Room-temperature ferroelectricity in CuInP2S6 ultrathin flakes. Nature Communications **7**, art. num. 12357 (2016).

[10] M. A. Susner, M. Chyasnavichyus, A. A. Puretzky, Q. He, B. S. Conner, Y. Ren, D. A. Cullen et al. Cation–Eutectic Transition via Sublattice Melting in $CuInP_2S_6/In_{4/3}P_2S_6$ van der Waals Layered Crystals. ACS Nano **11**, 7060 (2017).

[11] M. Osada, and T. Sasaki. The rise of 2D dielectrics/ferroelectrics. APL Materials **7**, 120902 (2019)

[12] C. Chen, H. Liu, Q. Lai, X. Mao, J. Fu, Z. Fu, and H. Zeng. Large-Scale Domain Engineering in Two-Dimensional Ferroelectric $CuInP_2S_6$ via Giant Flexoelectric Effect. Nano Letters **22**, 3275 (2022).

[13] V. Maisonneuve, V. B. Cajipe, A. Simon, R. Von Der Muhll, and J. Ravez. Ferrielectric ordering in lamellar $CuInP_2S_6$. Phys. Rev. B **56**, 10860 (1997).

[14] P. Toledano and M. Guennou. Theory of antiferroelectric phase transitions. Phys. Rev. B **94**, 014107 (2016)

[15] J. A. Brehm, S. M. Neumayer, L. Tao, A. O'Hara, M. Chyasnavichus, M. A. Susner, M. A. McGuire, S. V. Kalinin, S. Jesse, P. Ganesh, S. T. Pantelides, P. Maksymovych and N. Balke, Tunable quadruple-well ferroelectric van der Waals crystals. Nature Materials **19**, 43 (2020).

[16] W. Song, R. Fei, and L. Yang, Off-plane polarization ordering in metal chalcogen diphosphates from bulk to monolayer. Phys. Rev. B **96**, 235420 (2017).

[17] Yu. M. Vysochanskii, A. A. Molnar, M. I. Gurzan, V. B. Cajipe, and X. Bourdon. Dielectric measurement study of lamellar $CuInP_2Se_6$: successive transitions towards a ferroelectric state via an incommensurate phase? Sol. State Comm. **115**, 13 (2000).

[18] V. Liubachko, V. Shvalya, A. Oleaga, A. Salazar, A. Kohutych, A. Pogodin, and Yu. M. Vysochanskii. Anisotropic thermal properties and ferroelectric phase transitions in layered $CuInP_2S_6$ and $CuInP_2Se_6$ crystals. J. Phys. Chem. Sol. **111**, 324–327 (2017).

[19] A. Dziaugys, I. Zamaraite, J. Macutkevic, D. Jablonskas, S. Miga, J. Dec, Yu. Vysochanskii & J. Banys, Non-linear dielectric response of layered $CuInP_2S_6$ and $Cu_{0.9}Ag_{0.1}InP_2S_6$ crystals. Ferroelectrics **569**:1, 280, (2020).

[20] V. Samulionis, J. Banys, and Yu. Vysochanskii, Linear and Nonlinear Elastic Properties of $CuInP_2S_6$ Layered Crystals Under Polarization Reversal. Ferroelectrics **389**: 1, 18 (2009).

[21] J. Banys, J. Macutkevic, V. Samulionis, A. Brilingas & Yu. Vysochanskii, Dielectric and ultrasonic investigation of phase transition in $CuInP_2S_6$ crystals. Phase Transitions: A Multinational Journal **77**:4, 345 (2004).

[22] S. M. Neumayer, E. A. Eliseev, M. A. Susner, B. J. Rodriguez, S. Jesse, S. V. Kalinin, M. A. McGuire, A. N. Morozovska, P. Maksymovych and N. Balke. Giant negative electrostriction and dielectric tunability in a van der Waals layered ferroelectric. Phys. Rev. Materials **3**, 024401 (2019).

[23] V. Samulionis, J. Banys, and Yu. Vysochanskii, Piezoelectric and Ultrasonic Studies of Mixed $CuInP_2(S_XSe_{1-X})_6$ Layered Crystals. Ferroelectrics **351**:1, 88 (2007).

[24] Yu. Vysochanskii, R. Yevych, L. Beley, V. Stephanovich, V. Mytrovcij, O. Mykajlo, A. Molnar, and M. Gurzan. Phonon Spectra and Phase Transitions in $CuInP_2(Se_xS_{1-x})_6$ Ferroelectrics. Ferroelectrics **284**, 161 (2003).

[25] A. Dziaugys, K. Kelley, J. A. Brehm, Lei Tao, A. Puretzky, T. Feng, A. O'Hara, S. Neumayer, M. Chyasnavichyus, E. A. Eliseev, J. Banys, Y. Vysochanskii, F. Ye, B. C. Chakoumakos, M. A. Susner, M. A. McGuire, S. V. Kalinin, P. Ganesh, N. Balke, S. T. Pantelides, A. N. Morozovska, and P. Maksymovych. Piezoelectric domain walls in van der Waals antiferroelectric $CuInP_2Se_6$. Nature Communications **11**, Article number: 3623 (2020).

[26] A. N. Morozovska, E. A. Eliseev, K. Kelley, Yu. M. Vysochanskii, S. V. Kalinin, and P. Maksymovych. Phenomenological description of bright domain walls in ferroelectric-antiferroelectric layered chalcogenides. Phys. Rev. B **102**, 174108 (2020).

[27] A. N. Morozovska, E. A. Eliseev, S. V. Kalinin, Y. M. Vysochanskii, and Petro Maksymovych. Stress-Induced Phase Transitions in Nanoscale $CuInP_2S_6$. Phys. Rev. B **104**, 054102 (2021).

[28] E. A. Eliseev, Y. M. Fomichov, S. V. Kalinin, Yu. M. Vysochanskii, P. Maksymovich and A. N. Morozovska. Labyrinthine domains in ferroelectric nanoparticles: Manifestation of a gradient-induced morphological phase transition. Phys. Rev. B **98**, 054101 (2018).

[29] A. N. Morozovska, Y. M. Fomichov, P. Maksymovych, Yu. M. Vysochanskii, and E. A. Eliseev. Analytical description of domain morphology and phase diagrams of ferroelectric nanoparticles. Acta Materialia **160**, 109-120 (2018).

[30] A. N. Morozovska, E. A. Eliseev, Y. M. Fomichov, Yu. M. Vysochanskii, Victor Yu. Reshetnyak, and Dean R. Evans. Controlling the domain structure of ferroelectric nanoparticles using tunable shells. Acta Materialia **183**, 36-50 (2020).

[31] A. N. Morozovska, E. A. Eliseev, R. Hertel, Y.M. Fomichov, V. Tulaidan, V. Yu. Reshetnyak, and D. R. Evans. Electric Field Control of Three-Dimensional Vortex States in Core-Shell Ferroelectric Nanoparticles. Acta Materialia **200**, 256–273 (2020).

[32] E. A. Eliseev, A. N. Morozovska, R. Hertel, H. V. Shevliakova, Y. M. Fomichov, V. Yu. Reshetnyak, and D. R. Evans. Flexo-Elastic Control Factors of Domain Morphology in Core-Shell Ferroelectric Nanoparticles: Soft and Rigid Shells. Acta Materialia **212**, 116889 (2021).

[33] A. N. Morozovska, R. Hertel, S. Cherifi-Hertel, V. Yu. Reshetnyak, E. A. Eliseev, and D. R. Evans. Chiral Polarization Textures Induced by the Flexoelectric Effect in Ferroelectric Nanocylinders. Phys. Rev. B **104**, 054118 (2021).

[34] A. N. Morozovska, E. A. Eliseev, S. Cherifi-Hertel, D. R. Evans and R. Hertel. Electric field control of labyrinthine domains in core-shell ferroelectric nanoparticles (http://arxiv.org/abs/2207.00103)

[35] Yu. M. Vysochanskii, M.M. Mayor, V. M. Rizak, V. Yu. Slivka, and M. M. Khoma. The tricritical Lifshitz point on the phase diagram of $Sn_2P_2(Se_xS_{1-x})_6$. Soviet Journal of Experimental and Theoretical Physics **95**, 1355 (1989).

[36] A. Kohutych, R. Yevych, S. Perechinskii, V. Samulionis, J. Banys, and Yu. Vysochanskii. Sound behavior near the Lifshitz point in proper ferroelectrics. Phys. Rev. B **82**, 054101 (2010).

[37] A. N. Morozovska, M. D. Glinchuk, E. A. Eliseev. Phase transitions induced by confinement of ferroic nanoparticles. Phys. Rev. B **76**, 014102 (2007).

[38] A. N. Morozovska, I. S. Golovina, S. V. Lemishko, A. A. Andriiko, S. A. Khainakov, and E. A. Eliseev. Effect of Vegard strains on the extrinsic size effects in ferroelectric nanoparticles. Phys. Rev. B **90**, 214103 (2014).

[39] A. N. Morozovska, Y. M. Fomichov, P. Maksymovych, Y. M. Vysochanskii, and E. A. Eliseev. Analytical description of domain morphology and phase diagrams of ferroelectric nanoparticles. Acta Mater. **160**, 109 (2018).

[40] E. A. Eliseev, A. V. Semchenko, Y. M. Fomichov, M. D. Glinchuk, V. V. Sidsky, V. V. Kolos, Yu. M. Pleskachevsky, M. V. Silibin, N. V. Morozovsky, A. N. Morozovska. Surface and finite size effects impact on the phase diagrams, polar and dielectric properties of $(Sr,Bi)Ta_2O_9$ ferroelectric nanoparticles. J. Appl. Phys. **119**, 204104 (2016).

[41] L. D. Landau, E. M. Lifshitz, L. P. Pitaevskii. Electrodynamics of Continuous Media, (Second Edition, Butterworth-Heinemann, Oxford, 1984).

[42] A. K. Tagantsev and G. Gerra. Interface-induced phenomena in polarization response of ferroelectric thin films. J. Appl. Phys. **100**, 051607 (2006).

[43] P. Guranich, V. Shusta, E. Gerzanich, A. Slivka, I. Kuritsa, O. Gomonnai. "Influence of hydrostatic pressure on the dielectric properties of $CuInP_2S_6$ and $CuInP_2Se_6$ layered crystals." Journal of Physics: Conference Series **79**, 012009 (2007).

[44] A. V. Shusta, A. G. Slivka, V. M. Kedylich, P. P. Guranich, V. S. Shusta, E. I. Gerzanich, I. P. Prits, Effect of uniaxial pressure on dielectric properties of $CuInP_2S_6$ crystals. Scientific Bulletin of Uzhhorod University. Physical series, **28**, 44 (2010).

[45] A. Molnar, K. Glukhov, M. Medulych, D. Gal, H. Ban, Yu. Vysochanskii. The effect of changes in chemical composition and uniaxial compression on the phase transition of $CuInP_2S_6$ crystals, Abstract book of the FMNT 2020 Online Conference, Virtual Vilnius, Lithuania, 23 - 26 November (2020).

[46] V. Samulionis, J. Banys, Yu. Vysochanskii, and V. Cajipe. Elastic and electromechanical properties of new ferroelectric-semiconductor materials of $Sn_2P_2S_6$ family. Ferroelectrics **257:1**, 113 (2001).

[47] A. Kohutych, V. Liubachko, V. Hryts, Yu. Shiposh, M. Kundria, M. Medulych, K. Glukhov, R. Yevych, and Yu. Vysochanskii. Phonon spectra and phase transitions in van der Waals ferroics MM'$P_2X_6$, Molecular Crystals and Liquid Crystals (2022), https://doi.org/10.1080/15421406.2022.2066787

[48] Sumedha and S. Mukherjee, Emergence of a bicritical end point in the random-crystal-field Blume-Capel model, Phys. Rev. **E**, 101, 042125 (2020)

[49] https://www.wolfram.com/mathematica